\documentclass[a4paper,fleqn]{cas-dc}

\DeclareUnicodeCharacter{2212}{\ensuremath{-}}
\DeclareUnicodeCharacter{03C0}{\ensuremath{\pi}}

\usepackage[numbers]{natbib}
\setcitestyle{square,comma,numbers}

\def\tsc#1{\csdef{#1}{\textsc{\lowercase{#1}}\xspace}}
\tsc{WGM}
\tsc{QE}

\begin{document}
\let\WriteBookmarks\relax
\def\floatpagepagefraction{1}
\def\textpagefraction{.001}

\shorttitle{AgNP - \textit{Inula Viscosa} - kinetic}    

\shortauthors{E.Z. Okka, T. Tongur, T.T. Aytas, M. Yilmaz, {\"{O}}. Topel and R. Sahin}  

  \title[mode = title]{Green Synthesis and the formation kinetics of silver nanoparticles in aqueous \textit{Inula Viscosa} extract}



\address[1]{Department of Physics, Akdeniz University, 07058 Antalya, Turkey}
\address[2]{Department of Physics, Giresun University, Giresun, Turkey}
\address[3]{Department of Chemistry, Akdeniz University, 07058 Antalya, Turkey}
\address[4]{Necmettin Erbakan University Faculty of Engineering Department of Basic Sciences, 42090, Meram, Konya, Turkey}

\author[1,2]{Esra Zeybeko\u{g}lu Okka}[orcid=0000-0002-5427-9313]
\credit{Investigation, Methodology, Formal analysis, Writing - Original Draft}

\author[3]{Timur Tongur}[orcid=0000-0003-3030-8930]
\credit{Investigation, Formal analysis, Writing - Original Draft}

\author[1]{Taner Tarik Aytas}[orcid=0000-0003-4338-1117]
\credit{Investigation, Formal analysis}

\author[4]{Mucahit Yilmaz}[orcid=0000-0002-3387-5095]
\credit{Investigation, Formal analysis, Writing - Review and Editing}

\author[3]{{\"{O}}nder Topel}[orcid=0000-0003-1584-5519]
\credit{Conceptualization, Methodology, Writing - Review and Editing}

\author[1]{Ramazan Sahin}[orcid=0000-0002-9269-1528]
\credit{Conceptualization, Methodology, Writing - Review and Editing}
\cormark[1] 
\ead{rsahin@itu.edu.tr} 

\cortext[cor1]{Corresponding author} 




\begin{abstract}
In this study, we present the production of silver nanoparticles in aqueous \textit{Inula Viscosa} extract by the green synthesis approach at room temperature. The structural, morphological properties as well as formation kinetics of the synthesized silver nanoparticles were characterized by UV-VIS, STEM, XRD, Raman and FTIR measurements. Mono-dispersed and very stable silver nanoparticles with size of 15$\pm$5 nm and face-centered cubic crystal structure were synthesized in aqueous \textit{Inula Viscosa} extract. The kinetic studies of silver nanoparticles formation in \textit{Inula Viscosa} extract show that silver nanoparticle formation reaction reached the equilibrium within 24 h and fit in the first-order reaction kinetics. The results clearly show that the size of fabricated nanoparticles is independent on the dynamical formation process since the reaction time and initial silver ion concentration did not affect on size and morphology of the produced particles.
\end{abstract}


\begin{highlights}
\item Silver nanoparticles were synthesized by green synthesis method in \textit{Inula Viscosa} extract at ambient conditions.
\item The structural and morphological characterization of silver nanoparticles were determined.
\item The kinetic characteristic of the particle formation was also explored which causes more flexibility and control over fabrication processes.
\end{highlights}


\ExplSyntaxOn
\keys_set:nn { stm / mktitle } { nologo }
\ExplSyntaxOff

\maketitle

\section{Introduction}\label{introduction}

Fabrication of nanometer-sized particles (NPs) is a very popular and rapidly developing field for various branches of science such as physics, chemistry, biology and bioengineering \cite{qin_visual_2017,pandey_green_2012}. They are commonly used in daily life such as clothing, self-cleaning windows, hydrophobic-coated surfaces \cite{bhushan_nanomaterials_2020,lee_self-cleaning_2017,wigger_case_2017}. The nanoparticles ranging in size from 1 to 100 nm have many advantages due to confinement of the features such as thermal conductivity, chemical stability, optical-electrical and catalytic properties \cite{bhushan_nanomaterials_2020,sreekanth_green_2016} into a very small size as well as their larger surface area-volume ratios compared with their bulk form \cite{singh_green_2018,kanniah_green_2021,sarma_nanomaterials_2019}.

Among them, metal nanoparticles (especially Ag) are quite important for drug delivery because of their antimicrobial, antifungal and antibacterial effects \cite{bruna_silver_2021,prasher_silver_2018} and they are widely used in many technological and industrial areas such as optics, medicine, food safety and cosmetics \cite{duncan_applications_2011,ravichandran_nanotechnology_2010}. There are many techniques to fabricate metal nanoparticles that can be classified into; top-down and bottom-up processes \cite{zhu_synthesis_2020}. Although mechanical milling, chemical etching, sputtering and laser ablation can be given as an example in top-down method; chemical vapor deposition, sol-gel and spray pyrolysis can be listed for gathering molecules or atoms by chemical reactions for bottom-up approach \cite{noauthor_bottom-up_2004}.

Obtaining nanoparticles by physical and chemical methods have been studied for a long time \cite{iwahori_fabrication_2005,iravani_synthesis_nodate}. However, these methods require organic passivation agents to prevent nanoparticles from aggregation. Most of passivation agents are toxic and they pollute the environment in mass-production. Besides they require high cost and hazardous chemicals and special environmental conditions for synthesis such as temperature, pressure or pH. However, green synthesis of nanoparticles is cheap, straight-forward and less-toxic because of employing enviromentally friendly or toxic-free biological reagents such as plant, bacterial cell, algae and fungi for reduction from the metal salts \cite{smitha_green_2009,divya_biogenic_2019,edison_caulerpa_2016,stalin_dhas_facile_2014,metuku_biosynthesis_2014,bhargava_utilizing_2016,bonatto_higher_2014,chaudhari_dye_2022}. Plant extracts have attracted great interest in the green-synthesis of different types of nanoparticles compared to other bio-molecules due to their easy availability, biocompatibility, cost-effectiveness, and high stability \cite{chen_one-step_2012,karaagac_transfer_2020,karaagac_simultaneous_2020,ocsoy_aptamer-conjugated_2013}. The main components in a green synthesis are a plant extract of interest and the solution containing metal ion. The plant extract (such as mulberry leaves \cite{m_awwad_green_2012}) creates a medium to reduce of metal ions and provides a capping agent to stabilize the synthesized nanoparticles. Plant extracts mostly contain phenolic compounds, terpenoids, flavonoids, alkaloids and protiens which provide low-cost and non-hazardaous bioreduction of metal ions to metal nanoparticles (i.e. $Ag^{+}$ ions to $Ag^{0}$)  \cite{martinez-cabanas_antioxidant_2021}. 

Silver nanoparticles have been synthesized by many researchers using different plant extracts by green synthesis approach in the literature \cite{bar_green_2009,ahmad_green_2012,kumar_biosynthesis_2013,gardea-torresdey_alfalfa_2003}. In the studies, the researchers have mainly focused on the effects of experimental conditions on size and chemical composition of the fabricated silver nanoparticles. However, there is not much work to develop a detailed understanding about silver nanoparticle formation kinetics and mechanism in such complex plant extracts. A few studies were done on the kinetics on silver nanoparticle formation. Okafor et al. synthesized silver nanoparticles at 75 $^\circ$C and investigated  the reaction kinetics and biological activities \cite{okafor_green_2013}. Amini et al. fabricated silver nanoparticle with \textit{Avena Sativa} aqueous extract and hydro-alcoholic extract at different temperatures and concentrations of silver nitrate (AgNO$_{3}$) solution. The results showed that both diameter of the synthesized nanoparticles were inversely proportional to the temperature and the amount of nanoparticles depended on the AgNO$_{3}$ concentration.The synthesis conditions did not affect the size of nanoparticles \cite{amini_green_2017}. Prathna et al. tried to determine the temporal evolution of silver nanoparticles over time using experimental and theoretical models. Their results indicated that particles formation starts in as less as 2 hours reaction time. The agglomeration tendency increased at after the 4th hour of interaction. The particles conformed the Lifshitz-Slyozov-Wagner kinetics until 3.5 hours reaction time \cite{prathna_kinetic_2011}. In another work Hussain et al. carried out kinetic measurement of fabricated silver nanoparticles by green synthesis method. Their results showed that silver-mirror reaction provides an easy rout to prepare quantum dots. The formation kinetics of silver nanoparticles is independent from glucose concentration and the nucleation rate was proportional with ammonica \cite{hussain_time_2011}.

In the present study, we therefore investigated the formation kinetics of silver nanoparticles in \textit{Inula Viscosa} extract to control the particle formation mechanism and to develop a detailed understanding about silver nanoparticle fabrication in plant extracts. \textit{Inula Viscosa} is a self-growing species belonging to the Asteraceae family, which is common in the Mediterranean region. It contains many bioactive components such as guainolides, sesquiterpenes, lactones, flavonoids and essential oils \cite{ozkan_promising_2019}. Therefore, it does not only exhibit reducing and stabilizing properties during nanoparticle synthesis but also has high biological activities such as anti-inflammatory, anti-septic, antipyretic, antibacterial, anti-fungal \cite{gokbulut_antioxidant_2013,side_larbi_antibacterial_2016,talib_antiproliferative_2012}. In addition, \textit{Inula Viscosa} reduces the drug resistance of cancer cells so it can be preferred in phototherapy for supporting chemotherapy \cite{merghoub_inula_2016}.

Consequently, the synthesized silver nanoparticles in \textit{Inula Viscosa} extract are promising to have important biological activities such as antibacterial, anti-fungal, anti-cancer etc. Silver nanoparticles were synthesized by green synthesis approach using \textit{Inula Viscosa} extract at room temperature. The effects of the synthesis conditions such as concentration, reaction time and light on silver nanoparticle formation, particle size and morphology by means of UV-VIS spectrophotometry,  STEM microscopy as well as Raman and FTIR spectrometry. The formation kinetics of silver nanoparticles were also studied by spectrophotometric method using a UV-VIS spectrophotometer.

\subsection{Materials}
{\textit{Inula viscosa} plants were gathered from Manisa region, Turkey. The plant leaves were first separated from their branches, washed several times with deionized water, and then dried at room temperature. The washed and dried \textit{Inula Viscosa} leaves were stored in the laboratory under dark and dry conditions. AgNO$_{3}$ as silver (I) ion precursor was purchased from Sigma-Aldrich.
	
	\subsection{Preparation of silver nanoparticles using \textit{Inula Viscosa} extract} \label{sec:preparation}
	
	The plant leaves were put into the boiling (100$^\circ $C) deionized water and kept for 20 minutes. After cooling at room temperature, the \textit{Inula Viscosa} extract has been centrifuged at 12000 rpm for 10 minutes, filtered using Whatman No. 1 paper filter and then stored at 4$^\circ$C in dark for further studies. 
	
	The yellowish \textit{Inula Viscosa} extract was used as reducing and capping agent during nanoparticles fabrication. A certain amount of solid AgNO$_{3}$ was dissolved in deionized water and stirred with a magnetic stirrer to prepare 5 mM, 10 mM and 20 mM AgNO$_{3}$ solutions and the solutions were stored in dark until their usage. 0.75 mL of 5 mM AgNO$_{3}$ solution was gently mixed with 2.5 mL of \textit{Inula Viscosa} extract in a screw capped polypropylene (PP) tube. The same process was implemented to 10 mM and 20 mM AgNO$_{3}$ solutions to investigate the effect of AgNO$_{3}$ concentration on the silver nanoparticle formation. 
	
	The final silver (I) ion concentration as an initial concentration in the synthesis media after mixing \textit{Inula Viscosa} extract became 1.15, 2.30 and 4.60 mM, respectively. To investigate time-dependent silver nanoparticle formation with three different silver (I) ion concentration, ten parallel synthesis mixtures for each specific time were prepared and kept at 21-23 $^\circ$C under stirring and light conditions for the specified times (0.5, 1, 2, 3, 4, 5, 6, 8 and 24 hours). Silver nanoparticles formation was initially observed with change of the solution in color from light yellow to purple, reddish brown and dark brown depending on silver nanoparticles concentration (see Fig.\ref{fig:fig1}). To examine the effect of light-induced  nanoparticle formation, experiments were performed in the dark conditions with the same AgNO$_{3}$ concentrations and 24-hours reaction times. All reactions were started simultaneously and stopped as soon as the corresponding reaction time was completed. Each samples were centrifuged at 12000 rpm for 10 minutes and washed with deionized water for three times. The synthesized silver nanoparticles were finally dispersed in 3 mL of deionized water by lefting in a ultrasonic bath for five minutes. The purple silver nanoparticles dispersions were very stable and any sedimentation was not observed even after one month.
	
	\begin{figure}[h!]
		\centering
		\includegraphics[width=\linewidth]{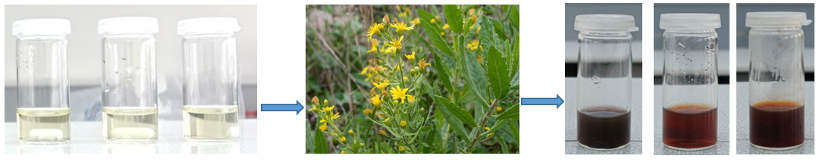}
		\caption{The synthesis of silver nanoparticles (after 24 h the colour changed to purple, reddish brown and dark brown)}
		\label{fig:fig1}
	\end{figure}

\subsection{Instrumentation}

All kinetic studies were carried out by Agilent Cary 100 UV-VIS double-beam spectrophotometer. The spectrophotometer was operated at “scan mode” between the spectral range of 350-800 nm for all measurements. The wavelength of maximum absorption ($\lambda_{max}$) of silver nanoparticles was obtained in between 450-480 nm according to the $\lambda_{max}$ versus time plot (see Fig.\ref{fig:fig2}). Amount of silver nanoparticles formed in the synthesis media was quantified using a Shimadzu Flame Atomic Absorption Spectrophotometer (flame-AAS) 7000F. The operating parameters of AAS-7000F are as follows; measurement wavelength of silver is 328.1 nm, slit width is 0.7 nm and air-acetylene gas flow is 2.2 min$^{-1}$. 

For the structural analysis such as shape, size, and morphology of the synthesized silver nanoparticles STEM images were taken with ZEISS GeminiSEM 500 at 200 KX magnification. Structural characterization of the silver nanoparticles produced at 24 hours reaction time for all concentrations was studied by using grazing incidence X-ray diffraction (GIXRD) analysis with CuK$_\alpha$  radition ($\lambda$=1.5406 {\AA}) on Panalytical Empyrean X-Ray Diffractometer in the range of 10$^\circ$-90$^\circ$  at a scan rate of 1.2$^\circ$ min$^{-1}$.

Raman analysis were carried out in the range of 200-1700  cm$^{-1}$ Renishaw inVia Confocal Raman Spectroscopy system. Raman data were collected by excitation of silver nanoparticles with a 3mW green light laser for 1 second periods and 250 accumulations in order to get information about molecular structure and vibrational properties. In addition, the FTIR measurements of silver nanoparticles were obtained using a Thermo Nicolet IS5 system in the region from 4000 to 600 cm$^{-1}$.
\subsection{Kinetic studies} 

Prior to progress kinetic studies, the linear working range of spectrophotometer for measurement of synthesized silver nanoparticles was determined. For this aim, the most concentrated AgNO$_{3}$ solution was selected. The sampling procedure about kinetic measurement is given in Section \ref{sec:preparation}. Then, the sample containing silver nanoparticles was diluted to different proportions and analyzed at spectrophotometer. After determining the linear working range to acquire calibration curve, kinetic studies were performed. As mentioned in Section \ref{sec:preparation}, plant extract and different initial concentrations of AgNO$_{3}$ were mixed at same sampling volume and they were stored at the same conditions for kinetic studies. All AgNO$_{3}$ added plant extract solutions were kept under sunlight for 0.5, 1, 2, 3, 4, 5, 6, 8 and 24 hours. Then, all samples were analyzed at spectrophotometer to follow the silver nanoparticle formation. 

\section{Results and Discussion}
\subsection{Formation and growth kinetics of silver nanoparticles in \textit{Inula Viscosa} extract}

Silver nanoparticles were synthesized by the green synthesis method using the \textit{Inula Viscosa} extract prepared by brewing for 20 minutes in boiled-water at room temperature under the ambient conditions. Besides, formation and growth kinetics of silver nanoparticles in the \textit{Inula Viscosa} extract were investigated  to explore the optimum synthesis conditions. For this, three different initial silver (I) ion concentrations  (1.15 mM, 2.30 mM and 4.60 mM) were employed for the same amount of \textit{Inula Viscosa} extract and time-dependent nanoparticle formation was acquired by a UV/VIS spectrometer in the spectral range of 350-800 nm. Fig. \ref{fig:fig2} shows the absorption spectra measured for above silver (I) ion concentrations. Characteristic surface plasmon resonance (SPR) band of silver nanoparticles between 400-500 nm is a unique evident for the presence of silver nanoparticles. Any shift in the peak position gives important information about the size and the formation mechanism of silver nanoparticles during a standard synthesis \cite{khan_optimization_2020,sanchez_leaf_2016}. From the absorption spectra in Fig.\ref{fig:fig2}, it can be seen that the peak positions stayed nearly unchanged within the standard deviation limits in time, especially for the higher initial silver (I) ion concentrations (See Figs.\ref{fig:fig2}a-d). The peak position at the higher concentrations (2.3 mM and 4.6 mM) keeps its spectral position constant as 450 nm through 24 hours (Fig. 2d). This means that silver nanoparticle formation regularly progresses with a significant size distribution under the concentrated silver (I) ion conditions. In the diluted synthesis conditions such as 1.15 mM, the peak position reached a constant value within 2 hours and stayed stable along 24 hours even if a broad variation in the spectrum was observed in the first four hours (see Fig. \ref{fig:fig2}a). However, increasing of silver nanoparticle concentration with progressing the reaction allows the nanoparticle formation reaction to proceed more regularly in the synthesis conditions with the higher concentrations (2.3 mM and 4.6 mM). Furthermore, the absorption spectrum was getting narrower in time (Fig. \ref{fig:fig2}b and \ref{fig:fig2}c). In addition, this is also observed by increasing initial silver (I) ion concentration. Therefore, this is a clear evident that higher silver ion concentration during the synthesis regulates the reduction reaction so that silver nanoparticles having same mean sizes can form in \textit{Inula Viscosa} extract.

\begin{figure}[h!]
	\centerline{
		\includegraphics[width=\linewidth]{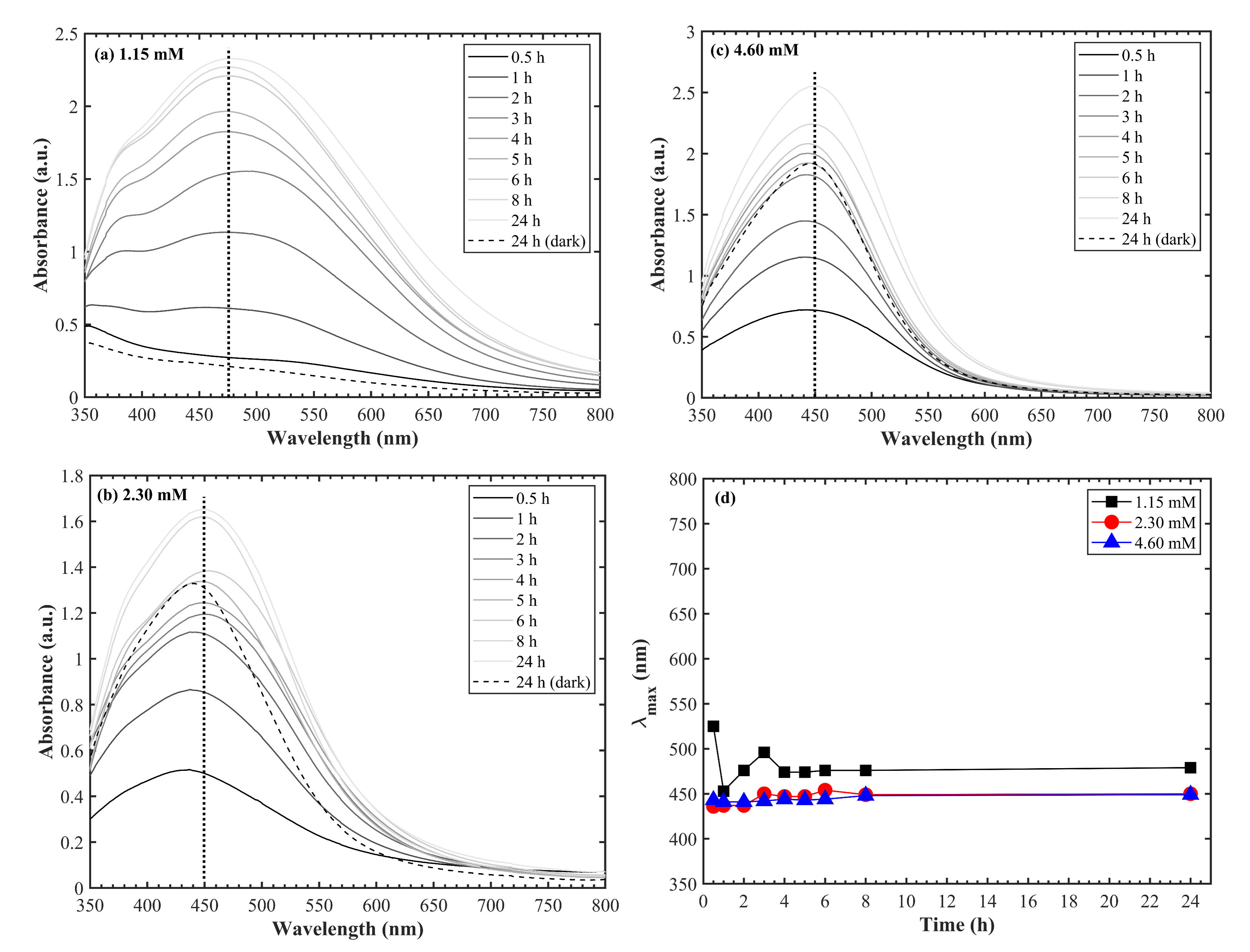}}
	\caption{The absorption spectra measured for three different initial silver (I) ion concentrations. (a) 1.15 mM, (b) 2.30 mM, (c) 4.60 mM and (d) the wavelength of maximum absorption }
	\label{fig:fig2}
\end{figure} 

The homogenous size distribution was also proved by STEM images taken in different reaction times at the concentration of 2.30 mM as seen in Fig.\ref{fig:fig3}. It is obvious that the size and morphology of the synthesized silver nanoparticles are almost identical from the beginning to the end of the formation process of silver nanoparticles in \textit{Inula Viscosa} extract (see Fig.\ref{fig:fig3} and Tab. \ref{tab:tab1}). The average size of the silver nanoparticles was determined as 15$\pm$5 nm and stayed approximately constant in time. Silver nanoparticles were also synthesized in dark conditions, however it was observed that the size and morphology of silver nanoparticles in \textit{Inula Viscosa} extract was not affected from the lightened and dark conditions. Dark conditions only affect the silver nanoparticle formation rate, and so the nanoparticle concentration in the reaction media (see the dashed line in Fig.\ref{fig:fig2}, Fig.\ref{fig:fig3}(h) and Tab.\ref{tab:tab1}). Consequently, the synthesis of silver nanoparticles in \textit{Inula Viscosa} extract is much dependent on the nature of the \textit{Inula Viscosa} extract. 

The previous study \cite{gokbulut_antioxidant_2013} shows that the \textit{Inula Viscosa} plant contains flavonoids such as quercetin, kaempferol, chlorogenic acid, rutin that a subgroup of polyphenols. It was known that flavonoids stabilize silver nanoparticles, yield homogeneous particle distribution and functioned as capping and reducing agent \cite{b_aziz_fabrication_2019,rafique_review_2017}. Therefore it can be said that the flavonoids of \textit{Inula Viscosa} extract ensured to obtain silver nanoparticles with a reaction of AgNO$_{3}$ solution. The effects of the plant extract ingredients to silver nanoparticles formation were analyzed in detailed by FTIR measurements to support this assumption. The results were given in Section \ref{characterization}.

\begin{figure}[h!]
	\centerline{
		\includegraphics[width=\linewidth]{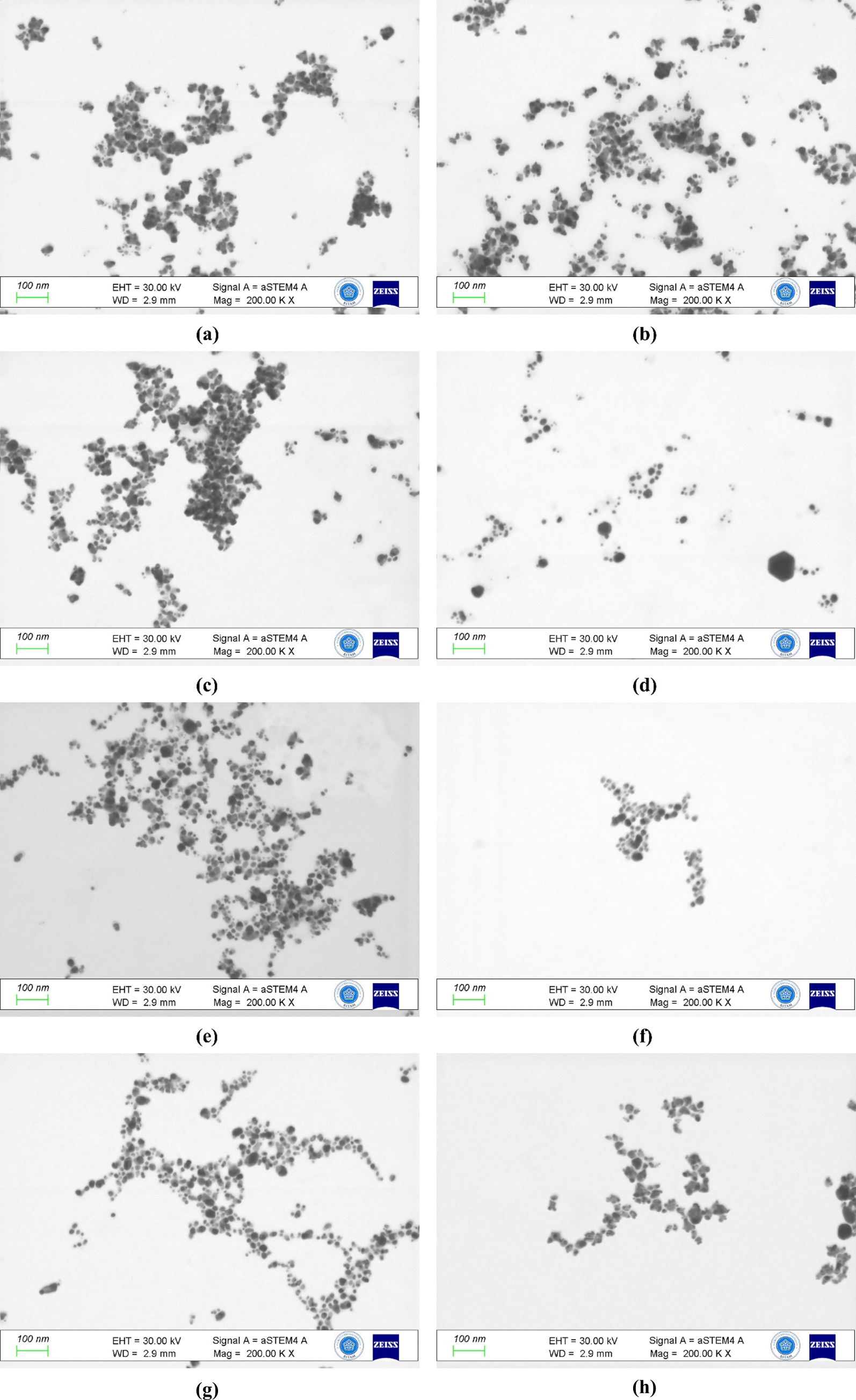}}
	\caption{STEM images of silver nanoparticles at the concentration of 2.30 mM. (a) 0.5 h, (b) 1 h, (c) 2 h, (d) 4 h, (e) 6 h, (f) 8 h, (g) 24 h and (h) 24 h dark}
	\label{fig:fig3}
\end{figure}

\begin{table}
	\caption{Particle size of silver nanoparticles at the concentration of 2.30 mM}
	\begin{tabular*}{\tblwidth}{@{} LLLL@{} }
		\toprule
		Time (h)  & Size (nm)\\
		\midrule
		0.5       & $12 \pm 4$  \\
		1         & $13 \pm 4$  \\
		2         & $14 \pm 5$  \\
		4         & $16 \pm 5$  \\
		6         & $17 \pm 5$  \\
		8         & $14 \pm 4$  \\
		24        & $15 \pm 5$  \\
		24 (dark) & $19 \pm 10$ \\
		\bottomrule
	\end{tabular*}
	\label{tab:tab1}
\end{table}


Growth kinetics of silver nanoparticles were also investigated based on the standard kinetic models to clarify the formation mechanism. Before kinetic analysis, the calibration curve was determined with UV-VIS spectrophotometer. It was found that absorption of silver nanoparticle indicates linear dependency between 0.23 to 2.4 absorption unit (see Fig.\ref{fig:Uv-Vıs}). After determining the linear range, the absorbance values that the point of calibration curve were converted to concentration (mg $L^{-1}$) by means of atomic absorption spectroscopy measurements to estimate silver nanoparticle concentration in the silver nanoparticle dispersions. For this purpose, 1 mL of silver nanoparticle dispersions being in the defined absorption range dried in oven at vacuum conditions. Then, the dried residues were dissolved with 0.2 mL pure nitric acid and diluted to 10 mL with deionized water. Thus, silver nanoparticles were converted to $Ag^{+}$ ions with applied treatment. Finally, concentrations of silver nanoparticles were calculated with the calibration curve that was obtained for $Ag^{+}$ with flame-AAS at 328.1 nm. The Coefficient of determination ($r^{2}$) of calibration curve was calculated as 0.9925. As a result, the absorption values of prepared samples given in Fig.\ref{fig:fig1} could be quantified and the concentration of silver nanoparticles depending on the reaction time could be obtained. Silver nanoparticle concentrations were also estimated by gravimetric method and the results appear within standard deviations. Fig. \ref{fig:fig5} shows how silver nanoparticle concentration changes depending on the reaction time and initial silver (I) ion concentration, and thus silver nanoparticle formation rate in \textit{Inula Viscosa} extract. In Fig. \ref{fig:fig5}, the nucleation and growth steps were observed similar to \cite{hussain_time_2011,litvin_kinetic_2012}.

\begin{figure}[h!]
	\centerline{
		\includegraphics[width=\linewidth]{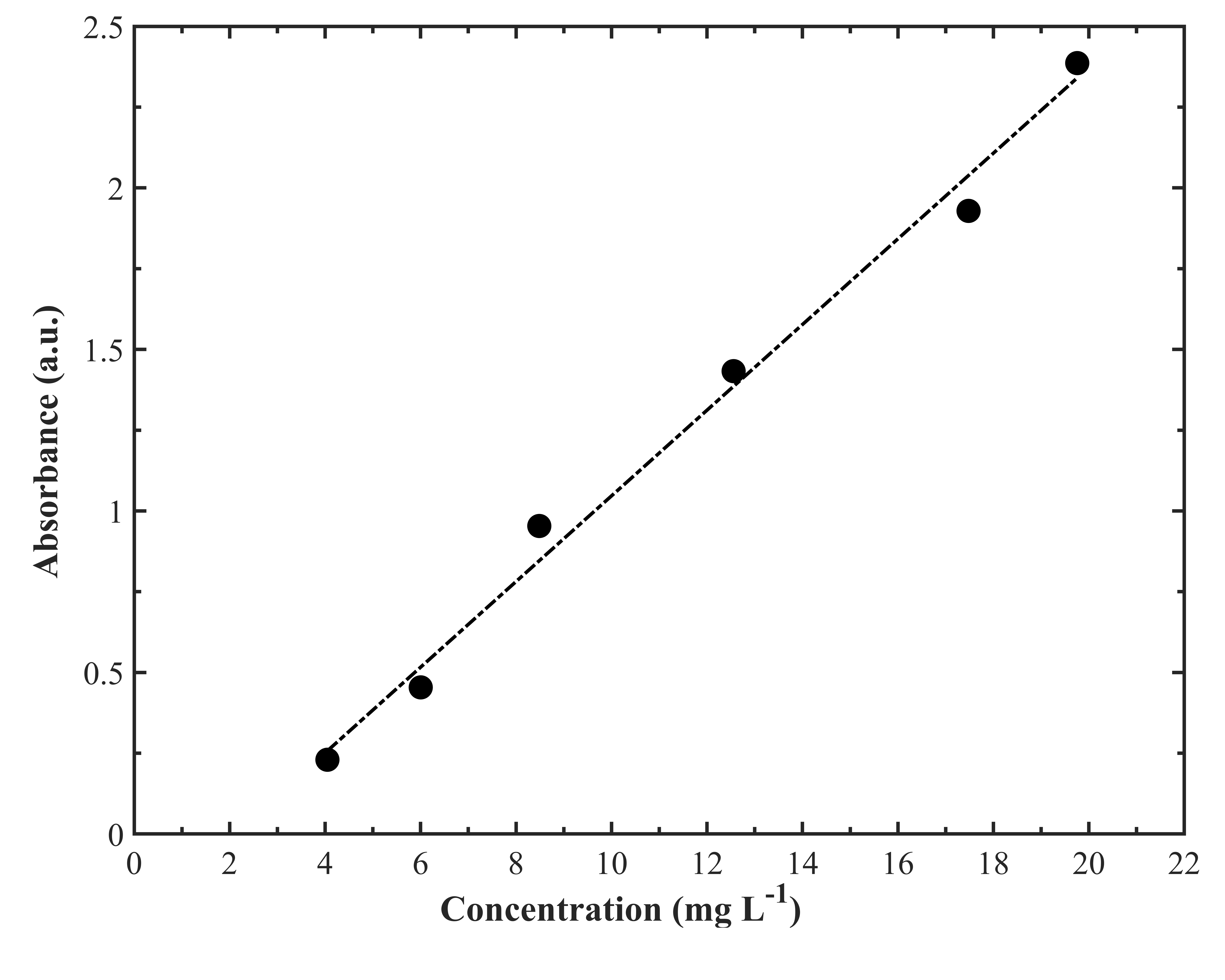}}
	\caption{Calibration curve of silver nanoparticles at UV-VIS spectrophotometer}
	\label{fig:Uv-Vıs}
\end{figure} 

Formation kinetics of silver nanoparticles were evaluated by fitting of experimental absorption spectra to first and second order kinetic models as explained follows. ln ($A_{\infty} - A_{0}$)/($A_{\infty} - A_{t}$) and 1/($A_{\infty} - A_{t}$) values were plotted as a function of time, respectively. Here, $A_{t}$ and $A_{\infty}$ are the absorption values at any time and at infinity, respectively. Rate of reactions were calculated from the slopes of ln ($A_{\infty}$ - $A_{0}$)/($A_{\infty}$ - $A_{t}$) versus time and 1/($A_{\infty}$ - $A_{t}$) versus time graphics. As a result, growth kinetics of silver nanoparticles were fitted to first-order kinetic model. ln ($A_{\infty}$ - $A_{0}$)/($A_{\infty}$ - $A_{t}$) versus time and 1/($A_{\infty}$ - $A_{t}$) versus time plots for 1.15 mM, 2.30 mM and 4.60 mM AgNO$_{3}$ were presented in Fig. \ref{fig:fig6}. The calculated coefficient of determinations ($r^{2}$) and rate of reactions for two kinetic model are given in Tab \ref{tab:tab2}.

\begin{figure}[htbp!]
	\centerline{
		\includegraphics[width=\linewidth]{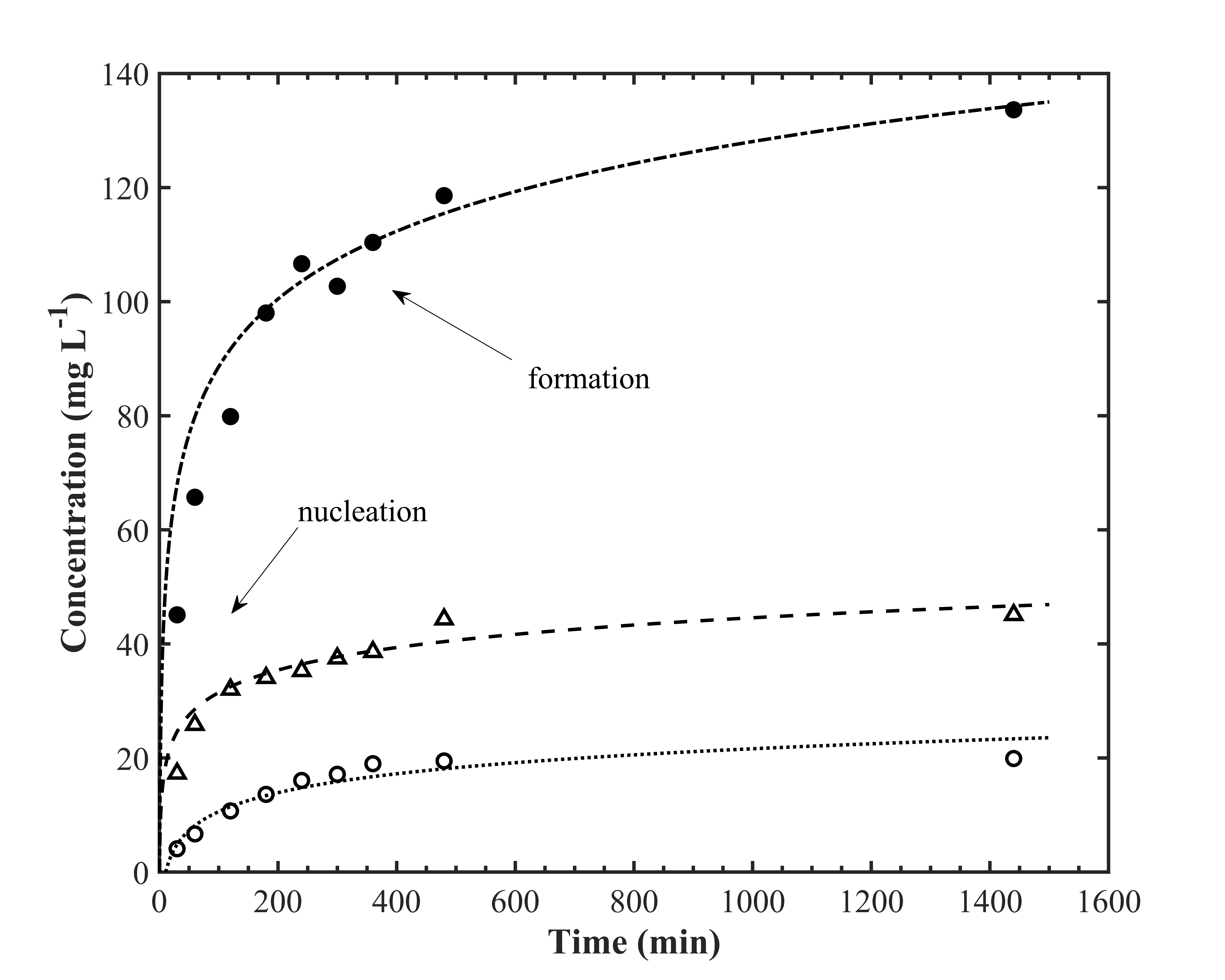}}
	\caption{The change of the concentration of silver nanoparticles depending on time. ($\circ$) 5 mM AgNO$_{3}$, ($\bigtriangleup$) 10 mM AgNO$_{3}$, ($\bullet$) 20 mM AgNO$_{3}$}
	\label{fig:fig5}
\end{figure}

\begin{figure}[htbp!]
	\centerline{
		\includegraphics[width=\linewidth]{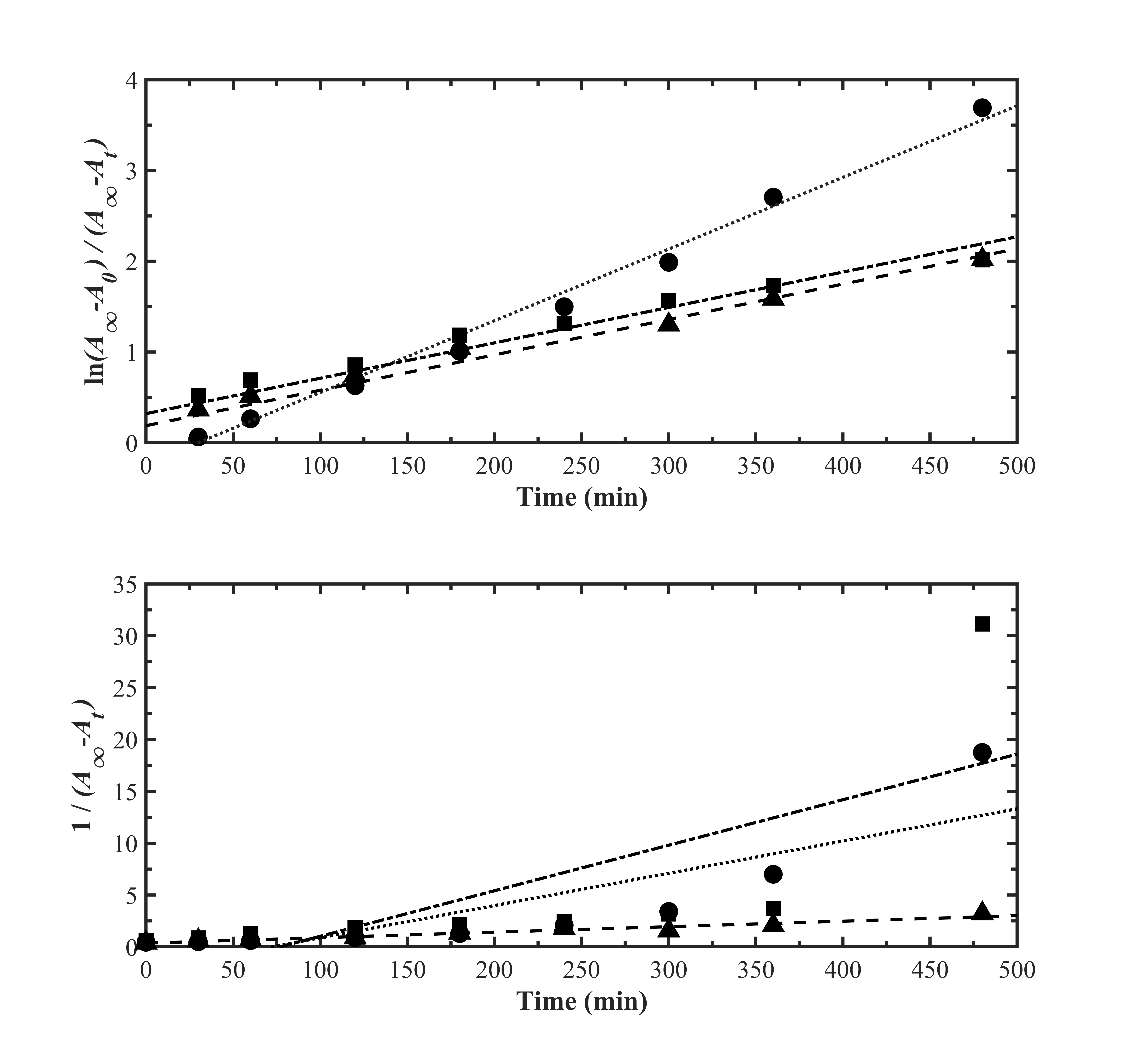}}
	\caption{\textbf{(a)} In $A_{\infty}-A_{0}$ vs time plots for first-order kinetic model. \textbf{(b)} $1/A_{\infty}-A_{t}$ vs time plots for second-order kinetic model. $(\bullet)$ 1.15 mM AgNO$_{3}$, $(\blacksquare)$ 2.30 mM AgNO$_{3}$, $(\blacktriangle)$ 4.60 mM AgNO$_{3}$. ($\bullet$,$\blacksquare$,$\blacktriangle$): experimental values, (- - -): best lines obtained by linear regression of experimental data with respect to the first and second order kinetic model}
	\label{fig:fig6}
\end{figure}

\begin{table}[htbp!]
	\caption{The coefficient of determinations ($r^2$) and rate of reactions for two kinetic models}
	\begin{tabular}{ccc|cc}
		\hline
		& \multicolumn{2}{c|}{first-order} & \multicolumn{2}{c}{second-order} \\ \cline{2-5} 
		& $r^2$            & $k(s^{-1})$                 & $r^2$            & $k$(L/mol min)                 \\
		1.15 mM & 0.9906           & $8.1 \times 10^{-3}$          & 0.7419           & $3.5 \times 10^{-2}$          \\
		2.30 mM & 0.9850           & $3.4 \times 10^{-3}$          & 0.5468           & $4.9 \times 10^{-2}$          \\
		4.60 mM & 0.9918           & $3.6 \times 10^{-3}$          & 0.9365           & $5.4 \times 10^{-3}$          \\ \hline
	\end{tabular}
	\label{tab:tab2}
\end{table}

\subsection{Characterization of silver nanoparticles synthesized in \textit{Inula Viscosa} extract }\label{characterization}

Systematic kinetic studies showed the nanoparticle formation reaction reached equilibrim within 24 hours. This time is therefore determined as the synthesis time for silver nanoparticles fabrication in \textit{Inula Viscosa} extracts and three parallel silver nanoparticles productions with different initial AgNO$_{3}$ concentrations as 1.15 mM, 2.30 mM and 4.60 mM were carried out under optimized reaction conditions. The structure properties of the synthesized silver nanoparticles such as their size,  morphology and surface modification were characterized by STEM, XRD, Raman spectroscpy and FTIR spectroscopy measurements.

The size and morphology of the synthesized silver nanoparticles in \textit{Inula Viscosa} extract were obtained using STEM  analysis. STEM images in Fig. \ref{fig:STEM} (a-c) clearly show that silver nanoparticle formation exhibits mono-dispersed behaviour in all initial concentrations. In addition, it can be seen from Fig. \ref{fig:STEM} (a-c) that the uniformity in the particle distribution increases with increasing the concentration. This situation was also observed in the UV/VIS spectra of silver nanoparticles where the surface plasmon resonance (SPR) bands get narrow with increasing the temperature. The SPR band of silver nanoparticles in UV/VIS spectra obtained for 24 h reaction time became sharper and shifted from 470 nm to 450 nm with the concentration increase (see Fig. \ref{fig:fig2} a-c). The sharp SPR band indicates the presence of spherical and small nanoparticles, which is also confirmed by STEM images in Fig. \ref{fig:STEM} \cite{peng_reversing_2010}. The shapes of silver nanoparticles become more spherical and the agglomerations disappear with increasing concentration (\textit{cf.} Fig. \ref{fig:STEM}).
Similarly, the presence of relatively larger and inhomogeneous particle distribution as seen in Fig \ref{fig:STEM}a come out as broader SPR band in UV/VIS spectra (Fig. \ref{fig:fig2}a). The average size of silver nanoparticles depending on concentration from STEM images are given in Fig. \ref{fig:STEM}d.} The results clearly show that the particle size of silver nanoparticles decreased from the most diluted conditions to more concentrated conditions; but, the particle sizes stayed identical within the experimental error limits under the concentrated conditions with 2.6 and 4.3 mM initial concentrations. This observation is also in aggrement with the results obtained from the absorption spectra in Fig. \ref{fig:fig2}.

\begin{figure}[h!]
\centerline{
	\includegraphics[width=0.9\linewidth]{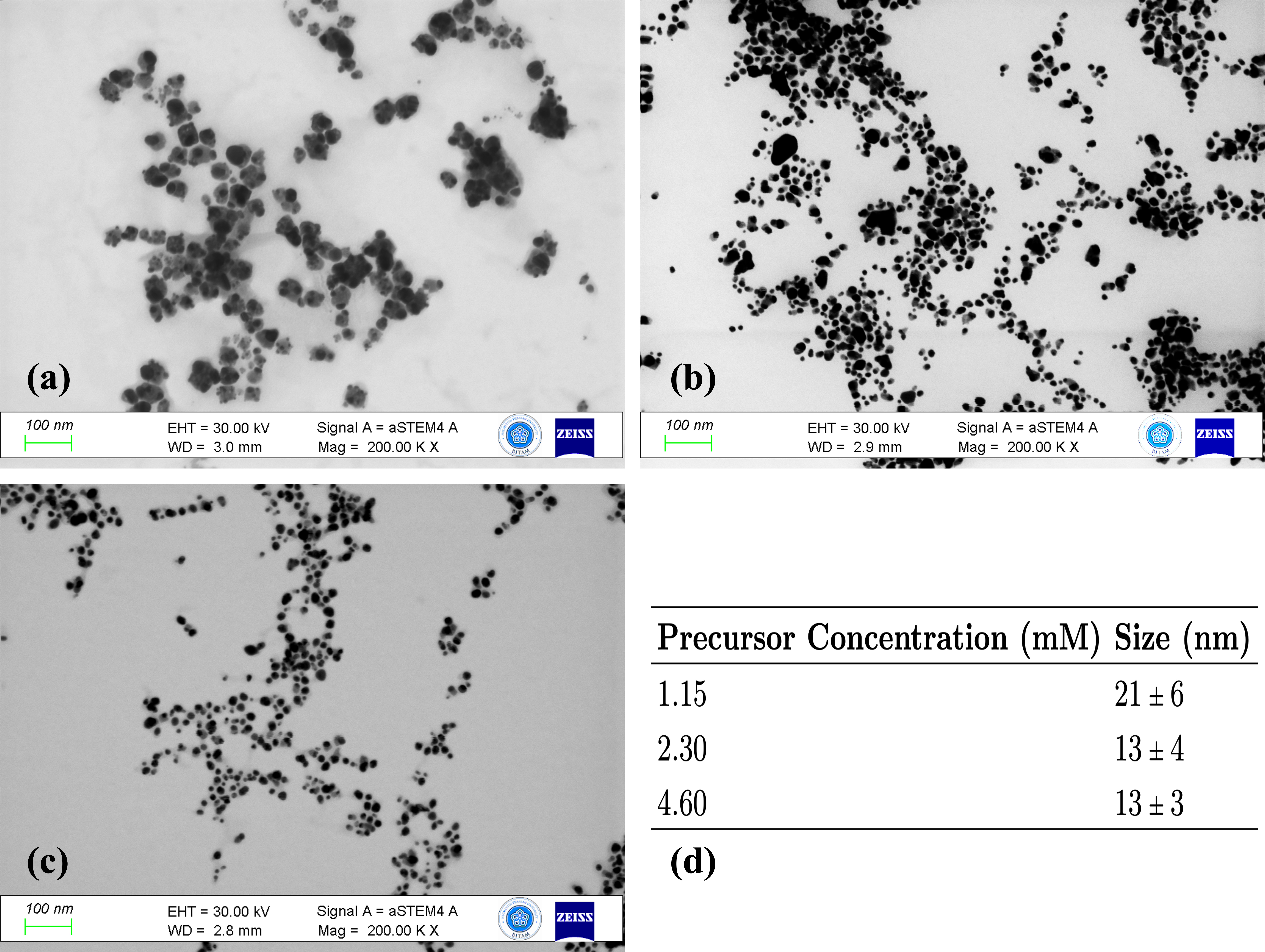}}
\caption{Comparative STEM images of silver nanoparticles synthesized with different initial concentrations at 24 h (a) 1.15 mM, (b) 2.30 mM, (c) 4.60 mM and (d) Average sizes of synthesized nanoparticles determined from the STEM images}
\label{fig:STEM}
\end{figure}

\begin{figure}[h!]
\centerline{
	\includegraphics[width=0.9\linewidth]{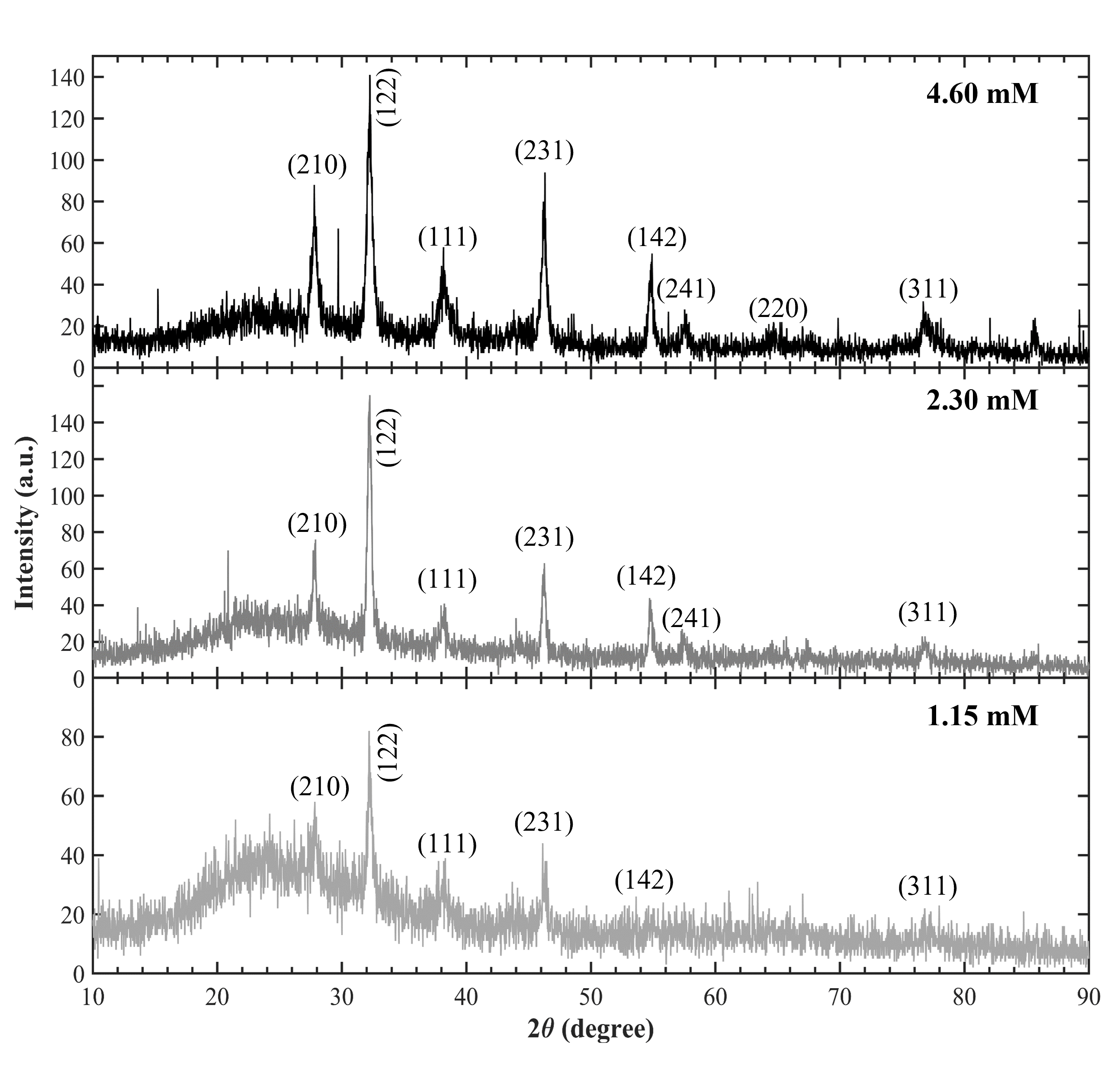}}
\caption{The XRD patterns of the silver nanoparticles for 1.15 mM, 2.30 mM, and 4.60 mM for 24 h reaction time}
\label{fig:XRD}
\end{figure}

Crystal phase and structural information of silver nanoparticles were obtained by XRD measurements. Bragg reflection peaks observed at 2$\theta$ values of   27.83$^\circ$, 32.19$^\circ$, 38.19$^\circ$, 46.23$^\circ$, 54.79$^\circ$, 57.49$^\circ$, 64.85$^\circ$ and 76.89$^\circ$ in the diffraction patterns were indexed to (210), (122), (111), (231), (142), (241), (220) and (311) planes on face-centred cubic (FCC) crystal structure with space group Fm3m (JCPDS file No. 04-0783) \cite{meng_sustainable_2015,priyadharshini_microwave-mediated_2014}. The reflections from (142), (241) and (220) planes could be detected as very weak peaks in diffraction pattern belonging the particles synthesize with 1.15 mM initial concentration due to the background noise at lower concentrations (Fig. \ref{fig:XRD}). Highly intense peak at 32.25$^\circ$ indexed to (122) suggests that majority of silver nanoparticles were formed with the same orientation (Fig. \ref{fig:XRD}). The other peaks with lower intensities can be attributed to the presence of bio-organic phases on the silver nanoparticle surfaces as explained before \cite{parameshwaran_green_2013}. All XRD results clearly show that silver nanoparticles were able to be synthesized with FCC crystal structure in \textit{Inula Viscosa} extract.

\begin{figure}[h!]
\centerline{
	\includegraphics[width=\linewidth]{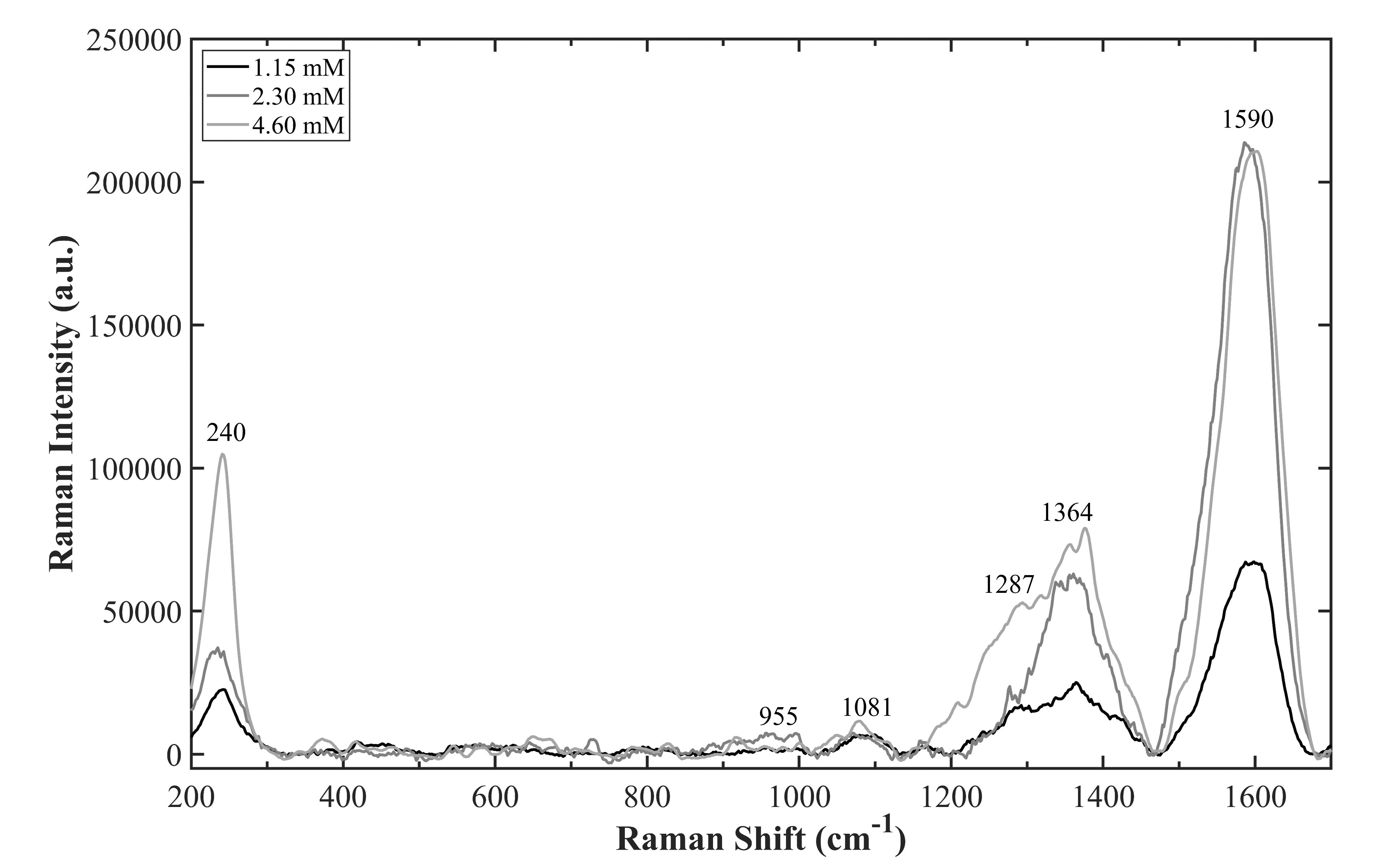}}
\caption{Raman spectra of silver nanoparticles prepared at the concentration of 1.15 mM, 2.30 mM and 4.60 mM }
\label{fig:Raman}
\end{figure}

\begin{figure}[h]
\centerline{
	\includegraphics[width=\linewidth]{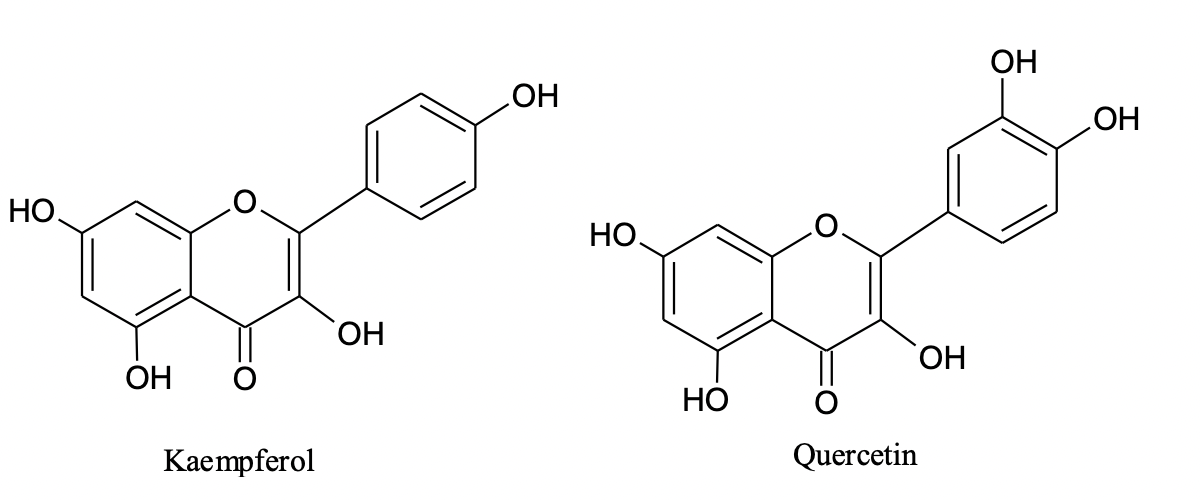}}
\caption{Structures of kaempferol and quercetin}
\label{fig:structure_of_keam}
\end{figure}

We also performed Raman spectrometric measurements to find out possible molecular interactions and chemical structure of silver nanoparticles as shown in Fig. \ref{fig:Raman}. Raman spectra shows distinctive vibrational modes at 240, 1287, 1364, and 1590 cm$^{-1}$ as well as  the peaks with lower intensity at 955 and 1081 cm$^{-1}$. \textit{Inula Viscosa} extract provides both reducing and stabilizing agents to obtain silver nanoparticles, and contains several organic compounds/components having hydroxyl, carbonyl, as well as amino groups. \cite{haoui_analysis_2015,kouache_experimental_2022}. The peak at 240 cm$^{-1}$ can be attributed to Ag-O or Ag-N stretching mode confirming the presence of the mentioned groups on the surface of silver nanoparticles \cite{chowdhury_concentration-dependent_2004}. In addition, the weak vibrational modes at 955 and 1081 cm$^{-1}$ can be related to C-C and C-N bonds, respectively \cite{podstawka_adsorption_2004}. Two sharp peaks at 1364 and 1590 cm$^{-1}$ correspond to symmetric and anti-symmetric C=O stretching vibration of carboxylic acid groups, respectively. The peak intensity in Raman spectra are proportional to the concentration. Any shift in the peak positions was not observed and they remained almost unchanged in their spectral positions (see Fig.\ref{fig:Raman}). It is well-known that \textit{Inula Viscosa} have many phenolic compounds \cite{ozkan_promising_2019,gokbulut_antioxidant_2013} and it was reported that quercetin and kaempferol are these of the most abundants in the aqueous \textit{Inula Viscosa} extracts. These flavonoids have carbonyl and hydroxyl groups having high reducing capacity, phenyl rings, heterocyclic ring containing the embedded oxygen atoms and conjugated double bonds in their structure (see Fig. \ref{fig:structure_of_keam}). \cite{kouache_experimental_2022,kheyar-kraouche_characterization_2018}.

\begin{figure}[h]
\centerline{
	\includegraphics[width=\linewidth]{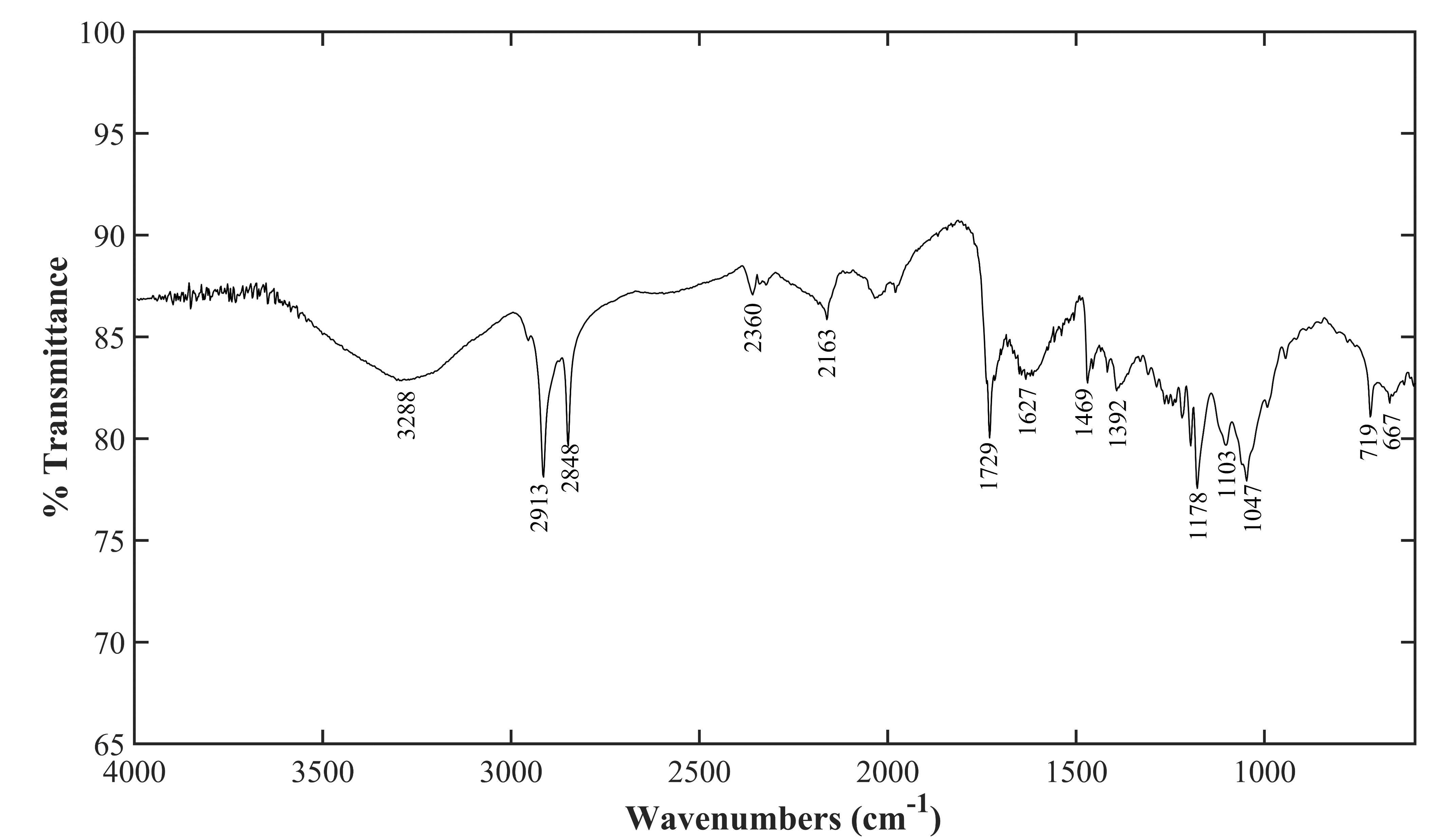}}
\caption{FTIR spectra of \textit{Inula Viscosa} extract containing reduced silver}
\label{fig:FTIR}
\end{figure}

The presence of the stabilizing compounds on the surface of the synthesized silver nanoparticles was also confirmed by FTIR measurements (see Fig. \ref{fig:FTIR}). The broad peak at 3288 cm$^{-1}$ in FTIR spectrum shows O-H bond stretching of hydroxyl groups on quercetin and kaempferol molecules. The peak at 1392 cm$^{-1}$ belongs to O-H bending vibrations of phenol. The peaks at 1178 cm$^{-1}$ to 1047 cm$^{-1}$ are also responsible for stretching vibrations of OH groups. The peaks observed at 2913 and 2848 cm$^{-1}$ belongs to C-H stretching vibrations. The presence of aromatic rings confirms with the peaks at 1627 cm$^{-1}$ and 1469 cm$^{-1}$ assigned to C=C stretching vibrations. The peak at 1729 cm$^{-1}$ comes from the C=O stretching vibration of carbonyl group. The peaks at 719 and 667 cm$^{-1}$ correspond to the aromatic C-H bending vibrations. Previous studies show that the functional groups such as hydroxyl, carbonyl, aldehyde, carboxylic acid groups of flavonoids responsible for the reduction and stabilization of silver nanoparticles \cite{hussain_applications_2019,ganaie_rapid_2016}. The FTIR results on the synthesized silver nanoparticles show clear evidence for the presence of the these reducing and stabilizing groups on the nanoparticle surface compared with the FTIR spectrum of pure quercetin \cite{noauthor_stability_2020}. Additionally, the FTIR spectrum of aqueous \textit{Inula Viscosa} extract in literature also proves similar streching and bending vibrations indicates the same functional groups \cite{kebir_valorization_2015}. On the light of the supposed mechanism for similar systems in literature, silver (I) ions in aqueous \textit{Inula Viscosa} extract are first bound to hydroxyl groups of flavonoids, and then reduced to metallic silver (Ag$^0$) resulting in silver nanoparticles \cite{noauthor_stability_2020}. After the reduction reaction, the molecules present in the medium stabilize silver nanoparticles most likely through the interaction of silver with $\pi$ bonds in their aromatic rings \cite{terenteva_formation_2015,b_aziz_fabrication_2019,maier_measurement_2015}. The synthesized silver nanoparticles in this study are easily dispersed in water and stay very stable long period of time which shows silver nanoparticles to be well-stabilized with capping molecules through the strong intercations, supported by FTIR measurements.

\section{Conclusions}

Green synthesis of silver nanomaterials is cheap and environmentally friendly approach and promising to have increased biological activities due to be fabricated in a synergistical synthesis medium containing anti-cancer, anti-bacterial and anti-fungal activities. Therefore in this study, silver nanoparticles were produced in aqueous \textit{Inula Viscosa} extract well-known to have high biological activities such as anti-inflammatory, anti-septic, antipyretic, antibacterial, anti-fungal as well as reduces the drug resistance reducing property in cancer treatment \cite{side_larbi_antibacterial_2016,talib_antiproliferative_2012}. Mono-dispersed and very stable silver nanoparticles with size of 15$\pm$5 nm and face-centered cubic crystal structure were synthesized in such a complex plant extract. The formation kinetics of silver nanoparticles in \textit{Inula Viscosa} extract were also investigated. It was not observed any effect of time and initial silver ion concentration on size and morphology of the produced particles. However, it was determined that silver nanoparticle formation reaction reached equilibrium within 24 hours and fit in the first-order reaction kinetics.








\section*{Acknowledgement}
We acknowledge the A.R.C. of Science and Technology (BITAM) for XRD and STEM measurements.
	
\printcredits


\bibliographystyle{elsarticle-num-names}


\end{document}